\documentclass[runningheads]{llncs}
\usepackage{graphicx}
\usepackage{booktabs} 
\usepackage{algorithm}
\usepackage{algpseudocode}
\usepackage{algorithmicx}
\usepackage{multirow}
\usepackage{diagbox}
\usepackage[caption=false]{subfig}
\usepackage{makecell}
\usepackage{tabularx}
\usepackage[hyphens]{url}

\newcommand{\repeatthanks}{\textsuperscript{\thefootnote}}

\newcommand\blfootnote[1]{%
  \begingroup
  \renewcommand\thefootnote{}\footnote{#1}%
  \addtocounter{footnote}{-1}%
  \endgroup
}

\begin{document}
\title{Taking Over the Stock Market: Adversarial Perturbations Against Algorithmic Traders}
\titlerunning{Adversarial Perturbations Against Algorithmic Traders}
\authorrunning{E. Nehemya* et al.}

\author{
    Elior Nehemya\thanks{The authors contributed equally}\orcidID{0000-0001-5363-3723} \and
    Yael Mathov\repeatthanks\orcidID{0000-0001-7004-1375} \and \\
    Asaf Shabtai\orcidID{0000-0003-0630-4059} \and
    Yuval Elovici\orcidID{0000-0002-9641-128X}
}
\institute{Department of Software and Information Systems Engineering, \\Ben-Gurion University of the Negev, Beer-Sheva 8410501, Israel \\
\email{$\{$nehemya,yaelmath$\}$@post.bgu.ac.il}\\
\email{$\{$shabtaia,elovici$\}$@bgu.ac.il}}

\maketitle              
\begin{abstract}
In recent years, machine learning has become prevalent in numerous tasks, including algorithmic trading.
Stock market traders utilize machine learning models to predict the market's behavior and execute an investment strategy accordingly.
However, machine learning models have been shown to be susceptible to input manipulations called adversarial examples.
Despite this risk, the trading domain remains largely unexplored in the context of adversarial learning.
In this study, we present a realistic scenario in which an attacker influences algorithmic trading systems by using adversarial learning techniques to manipulate the input data stream in real time.
The attacker creates a universal perturbation that is agnostic to the target model and time of use, which, when added to the input stream, remains imperceptible.
We evaluate our attack on a real-world market data stream and target three different trading algorithms.
We show that when added to the input stream, our perturbation can fool the trading algorithms at future unseen data points, in both white-box and black-box settings.
Finally, we present various mitigation methods and discuss their limitations, which stem from the algorithmic trading domain.
We believe that these findings should serve as an alert to the finance community about the threats in this area and promote further research on the risks associated with using automated learning models in the trading domain.

\end{abstract}

\section{Introduction}
\blfootnote{Accepted to ECML PKDD 2021 \url{https://2021.ecmlpkdd.org/wp-content/uploads/2021/07/sub_386.pdf}}
In recent history, stock markets have been a major vehicle for personal and institutional investing. 
When buying or selling financial assets via the stock exchange, traders gain or lose money based on changes in the assets' value.
To maximize his/her profits, the trader needs to accurately predict changes in the market.
Yet, predicting the prices of financial assets is a challenging task, due to the complex dynamics of the market's behavior.
To do so, a trader processes a vast amount of market data and uses it to try to predict the value of a specific stock.
Based on this prediction, the trader develops an investment strategy concerning the stock, which results in one of the following actions: 
(1) selling the stock, (2) buying more of the stock, or (3) holding his/her assets intact.

To gain a possible advantage, traders use computers to analyze financial market data quickly and execute trades automatically; known as algorithmic trading (AT), this is the most common form of stock trading performed today~\cite{treleaven2013algorithmic}.
Most AT systems use a similar process, consisting of three main steps: data preprocessing, applying prediction model, and investment strategy execution. 
The preprocessed data is used to predict the market's behavior using a prediction unit known as the alpha model, and based on this prediction, the best investment strategy is chosen and executed.
In the past, a set of predefined rules was used to calculate the market predictions and choose the investment strategy~\cite{coutts2000trading}, but today, the popularity of machine learning is changing the picture.
More traders are relying on machine learning-based alpha models, such as support vector machine (SVM) and artificial neural network (ANN) architectures~\cite{kumar2020stock}, which can automatically make multiple predictions in milliseconds based a large amount of data; the edge they provide is especially useful in high-frequency trading (HFT).
However, since most of the methods used to choose the investment strategy are still rule-based, the alpha model has become the heart of AT systems. 

In recent years, hackers have profited from the stock market by spreading fake news~\cite{fisher2013syrian} or published stolen sensitive information~\cite{domm2013false,bianchi2019cyber} on companies as a means of decreasing their stock price.
Additionally, cyber attacks can be used to compromise various players in the market so as to directly affect the market (e.g., by hacking traders' accounts and performing transactions on their behalf~\cite{nakashima2007hack,rooney2020hackers}).
Since no one is safe from those threats~\cite{neyret2020stock}, regulatory authorities fight them by monitoring the stock market to identify fraudulent activity performed by malicious or compromised entities. 
However, the nature of AT makes it challenging to monitor and understand the bots' behaviors, especially in HFT systems that perform many transactions in a short period of time, making it difficult to identify real-time changes in their operation.
Therefore, changes regarding the behavior of an AT bot can only be identified retrospectively, which may be too late.
Moreover, AT bots can use complex learning algorithms (i.e., ANNs), which remain a focus of research due to their lack of explainability~\cite{gunning2017explainable}.

Along with the potential cyber attacks and the monitoring challenges mentioned above, a new threat has emerged from the rapid technological improvements seen in recent years: 
Machine learning models have been shown to be vulnerable to adversarial inputs known as adversarial examples which are maliciously modified data samples that are designed so that they will be misclassified by the target model~\cite{szegedy2013intriguing}. 
This vulnerability threatens the reliability of machine learning models and could potentially jeopardize sensitive applications, such as those used in the trading domain.
By exploiting the existence of adversarial examples, an attacker can gain control of an AT bot's alpha model and, as a result, influence the system's actions.
Moreover, by gaining control of multiple AT systems, an attacker could put the entire stock market at risk.
Yet, adversarial perturbations in AT are rare, largely due to the data, which is extremely dynamic, unpredictable, and heavily monitored by law enforcement agencies;
thus, state-of-the-art methods successfully used in other domains may be ineffective in the trading realm.
Unlike images, where the features are pixels that can be easily perturbed, AT data is extracted from a live stream of the stock market, with numeric features that are inclined to change rapidly and sometimes in a chaotic manner.
Additionally, since AT alpha models use real-time data, crafting a perturbation for a specific data point in advance is challenging.
Due to the frequent changes in the data, by the time the attacker has crafted the adversarial perturbation, the data may have changed completely.
Since the attacker cannot predict the market's behavior, he/she cannot craft a perturbation in real time.

In this study, we investigate the existence of adversarial perturbations in the AT domain, taking all of the aforementioned challenges into account. 
We present a realistic scenario where an attacker that manipulates the HFT data stream can gain control of an AT bot's actions in real time.
To achieve this goal, we present an algorithm that utilizes known market data to craft a targeted universal adversarial perturbation (TUAP), which can fool the alpha model.
The algorithm is designed to create a small, imperceptible TUAP so as to avoid detection; the TUAP created is agnostic to the target alpha model, as well as to the unseen data samples to which it is added.
Our method is evaluated using real-world stock data and three different prediction models, in both white-box and black-box settings.
We also demonstrate different mitigation methods against our attack and discuss their limitations when used to protect AT systems.
Our results suggest that the risk of adversarial examples in the trading domain is significantly higher than expected.
Based on our review of the literature, it seems that adversarial learning in the trading domain largely remains an unexplored field of research.
Therefore, we encourage both regulators and traders to address this concern and implement the methods needed to reduce the risk caused by the use of vulnerable models for algorithmic trading.

\section{Background}

\subsection{Algorithmic Trading}
Algorithmic trading refers to the use of computer programs which perform trading transactions based on an analysis of the stock market. 
Both human and AT system traders aim to maximize their profit by perfecting their ability to predict a stock's future price, use it to define an investment strategy, and perform beneficial transactions~\cite{treleaven2013algorithmic}.
Since an accurate prediction results in a more profitable investment, the AT system maintains an alpha model that models the market's behavior.
The system also has an execution logic unit which turns the prediction into a transaction based on risk management policies.
A popular type of AT is HFT, where traders perform a large number of transactions in a short period of time~\cite{treleaven2013algorithmic}.
Since HFT requires split-second decisions, it relies solely on automated software~\cite{bigiotti2018optimizing}, and thus, we focus on this type of trading in this paper. 

To predict the future stock price, the alpha model obtains data from an online broker or other external sources.
There are two main types of features used for stock market prediction~\cite{kumar2020stock}: fundamental indicators and technical indicators, which are used for fundamental and technical analysis respectively. 
Fundamental analysis focuses on macro factors that might correlate with the stock's price, such as financial records, economic reports, and balance sheets. 
Conversely, technical analysis assumes that all of the relevant information is factored into the stock's price. 
More than 80\% of alpha models use technical indicators as input features~\cite{kumar2020stock}. 
Since the AT system's decisions are based on the market prediction, traders are constantly seeking new methods to improve their alpha models.
In the past, traders made predictions by manually building trading strategies based on known patterns in the market data stream~\cite{coutts2000trading}. 
However, increases in computational capabilities caused traders to switch to sophisticated statistical methods, which were later replaced by machine learning-based alpha models that were shown to better estimate the market's behavior.
The emergence of the big data era introduced new models, and traders requiring rapid analysis of the massive amount of market data began to use ANNs~\cite{giacomel2015algorithmic}, such as deep neural networks (DNNs)~\cite{arevalo2016high,chong2017deep} and recurrent neural networks (RNNs)~\cite{chenapplication,pang2020innovative}.

\subsection{Adversarial Learning}
The term adversarial example was first used in~\cite{szegedy2013intriguing} to describe a well-crafted perturbation added to an input sample that fools a target DNN. 
Crafting an adversarial example is done by adding an imperceptible perturbation to an input sample which results in misclassification by the model. 
Additionally, adversarial perturbations can fool other models, even those trained on different training sets, which is known as the transferability property of adversarial examples~\cite{szegedy2013intriguing}.
Those findings caused the research community to delve deeper in order to both better understand this vulnerability and develop new methods for crafting adversarial perturbations~\cite{goodfellow2014explaining,kurakin2016adversarial,carlini2017towards}.
To use adversarial examples in more realistic scenarios, some studies utilized the transferability property to perform an attack in black-box settings.
The attacker creates adversarial examples for a surrogate model and then transfers them to the target model, which he/she knows little about.
Initially, adversarial perturbations were crafted based on a specific data sample; this changed when the universal adversarial perturbation (UAP) was presented.
The UAP is a single perturbation that is able to fool a learning model when added to both the samples in the training set and unseen data samples, and is also transferable to other neural networks.
Since the attack allows an attacker to craft one perturbation and use it against unseen samples, it can be used in domains where the data is unknown (e.g., AT).

Initial research targeted images, but recent studies have expanded to other domains, yet the AT domain remained unexplored.
While the simple attacks targeting AT presented in~\cite{arnoldi2016computer} can easily be identified by regulation authorities, they demonstrate how AT increases volatility in the market.
Therefore, it is reasonable to suspect that adversarial examples could be used in the trading domain, since it increasingly relies on machine learning models.
Adversarial learning-based attack was demonstrated with a UAP that was used to identify transactions that manipulate the limit order book data and as a result cause the target AT bot to change its behavior~\cite{goldblum2020adversarial}. 
However, this method is limited to stocks with low trading volume, because by the time the attacker finalizes the transactions, the limit order book can completely change, which can make the malicious transactions less effective or not effective at all.

\section{Problem Description}

\subsection{Trading Setup}
\begin{figure}[t]
    \centering
    \includegraphics[width=0.8\textwidth]{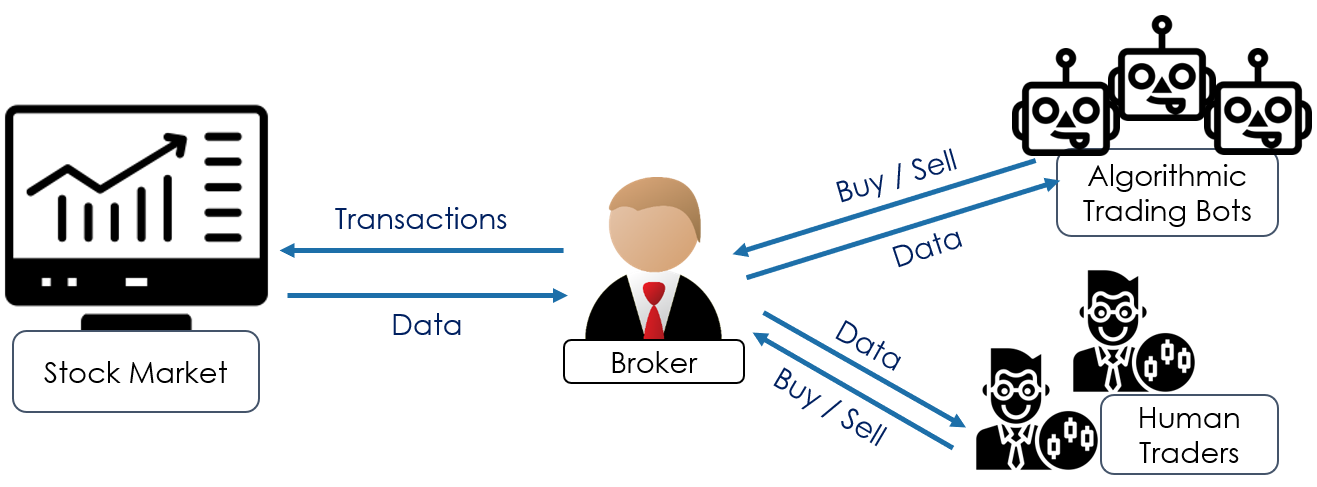}
    \caption{
    Simplified illustration of the HFT ecosystem. 
    All traders (humans and AT bots) collect market data and send transaction requests to the broker. 
    The broker executes the transaction and sends the market data to the trader.
    }
    \label{fig:trading_world}
\end{figure}

We assume the simplified HFT environment presented in Fig.~\ref{fig:trading_world}, with the following entities: a broker, traders (humans or AT bots), and the stock market.
Stock market transactions are limited to trusted members only, which may limit the traders' ability to exchange assets in the market~\cite{investopedia2019broker}.
Therefore, a broker is a trusted entity that connects traders and the stock market by receiving and executing buy and sell requests on behalf of the traders. 
Each transaction changes the market's supply and demand, thus affecting the stock price. 
Information about the market changes, including changes in stock prices, is sent to the traders via the broker.
However, data anomalies are common in HFT due to software bugs, errors in transaction requests, environmental conditions, and more~\cite{Traders2019Baker,nasdaqError}.
Thus, some AT systems embed anomaly detection filters during preprocessing and ignore abnormal changes in the data.
In this work, we focus on discount brokers who play a major role in HFT by providing online trading platforms to encourage frequent trade execution. 
The target AT system receives data from the broker, in the form of one-minute intraday stock prices, and processes and feeds it to a machine learning-based alpha model, which tries to predict whether the stock price will increase or decrease.
Based on the prediction, the AT system chooses an investment strategy and performs the corresponding action. 

\subsection{Threat Model}
\begin{figure}[t]
    \centering
    \includegraphics[width=0.9\textwidth]{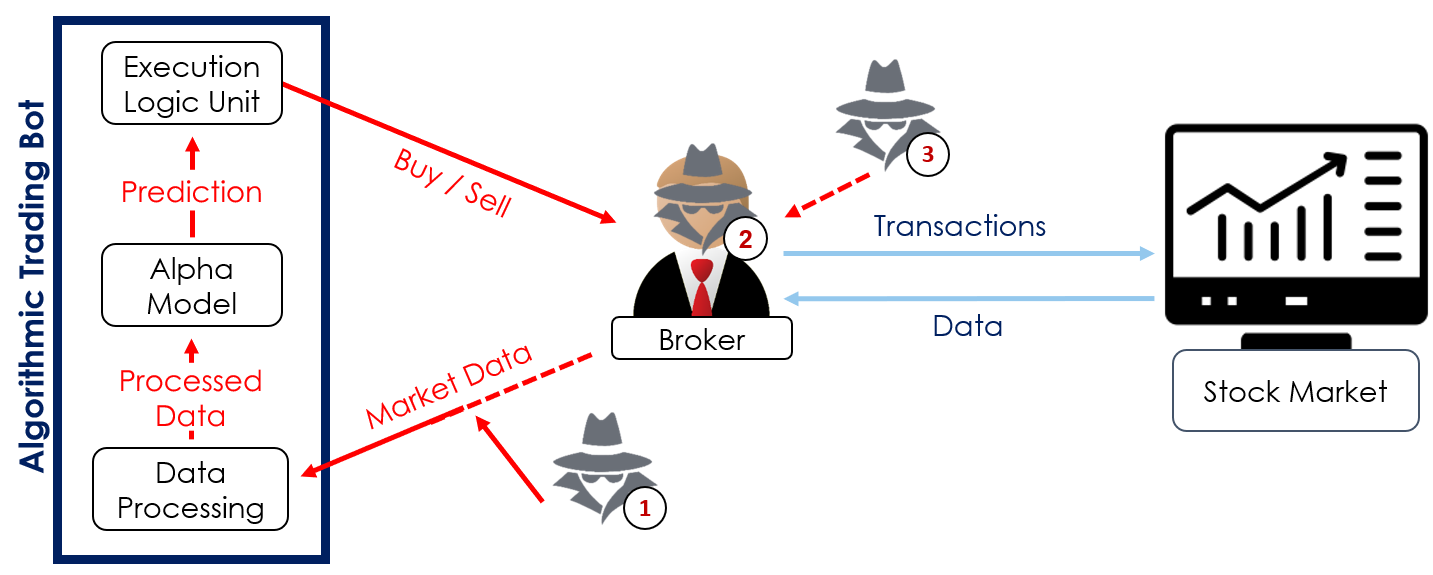}
    \caption{An illustration of the attack flow. 
    The attacker can be one of the following: 
    (1) A cyber attacker that targets a specific AT bot,
    (2) a malicious broker, or (3) a compromised broker.
    While (1) manipulates just the data feed for one AT bot, (2) and (3) perturb the data to all traders that collect data from the broker.
    }
    \label{fig:attack}
\end{figure}

Since fraudulent behavior or cyber crimes performed by any entity in the stock market can result in significant financial gain~\cite{neyret2020stock}, we can assume that even a broker cannot be trusted. 
Therefore, we consider the three threat models shown in Fig.~\ref{fig:attack}:
a cyber attacker performs a man-in-the-middle attack on the data stream sent from the broker to the bot and gains control of the data sent to the alpha model; a malicious broker manipulates the data he/she sends to the traders to personally benefit from the effects of the perturbed data on the traders; and a compromised broker unknowingly sends the traders data that was manipulated by an attacker.
In all cases, the attacker's goal is to profit financially or personally by sabotaging one or more AT systems.
We assume that the attacker can manipulate the market stream and send it to the AT system, which uses the data as an input.
Additionally, the attacker is aware of the existence of regulatory monitoring and the possibility that an anomaly detector might be used to filter major changes in the data and wants to bypass both by performing minor changes in the data stream.
We start by assuming that the attacker has complete knowledge of the target model, and later this assumption will be dropped.

\section{\label{sec:method}Proposed Attack}
As illustrated in Fig.~\ref{fig:attack}, the attacker wants to perturb the market data stream to fool the alpha model and control the AT system's behavior.
However, the trading domain introduces new challenges that need to be addressed to build a realistic attack against AT systems.
First, stock market data rapidly changes over time, and by the time the attacker crafts the perturbation, it may be no longer relevant, since the prices have likely changed.
Second, the attacker does not know the true label ahead of time, and perturbing the data might create an undesired outcome; if at a specific time both the target and true labels are 'increase,' adding a perturbation might cause an undesired effect in which the sample is classified as 'decrease.'
To address those challenges, we suggest a targeted version of a UAP~\cite{moosavi2017universal}. 
Since the TUAP is effective when added to unseen data, the attacker can craft one TUAP in advance and apply it to the data stream in real time. 
We craft the TUAP using past data samples, with equal representation of all classes, to ensure that input samples that were originally classified as the target class will output the same result after applying the TUAP.

We define $f$ to be the AT system's alpha model, which receives a snapshot $x$ of a stock's price for a period of $k$ minutes and outputs an estimation $f(x)$ for the price behavior in the next few minutes (increase or decrease). 
For a set of snapshots of stock prices $X$, a target label $y$, and an alpha model $f$, the attacker aims to craft TUAP $v$, such that the number of $x \in X$, where $f(x + v) = y$, is maximal.
Therefore, we define a successful attack using the targeted fooling rate (\emph{TFR}): the \emph{percentage} of $x \in X$ such that $f(x + v) = y$.
To maintain the imperceptibility and make the TUAP look like a normal price fluctuation, the size of $v$, $\|v\|_{2}$, should be minimal.
Thus, the attacker chooses two thresholds that determine the attack's requirements: $\delta$ denotes the minimal TFR value for $f(x + v)$ with regard to $y$, while $\epsilon$ defines the maximal perturbation size of $v$.

To craft the TUAP, Algorithm~\ref{attackAlgo} receives a set of data points $X$, a target label $y$, a classifier $f$, a minimal TFR $\delta$ threshold, and a maximal perturbation size $\epsilon$, and iteratively crafts a TUAP $v$ such that $TFR(f(X + v), y) \geq \delta$ and $\|v\|_{2} \leq \epsilon$. 
Thus, the TUAP is smaller than $\epsilon$, yet it fools $f$ for at least $\delta\%$ of $X$. 
To calculate the TUAP, we initialize $v$ to a zero vector (line 1), and split $X$ into batches to improve the convergence time of the optimization process. 
Next, we iteratively calculate TUAP $v$ by checking each batch $X_{B}$ in $X$, and if $TFR(f(X_B + v), y) \geq \delta$ (line 4), we update $v$ as follows:
First, we compute a perturbation $v_{i}$ with a minimal size and maximal TFR (line 5) using a modified version of a targeted BIM~\cite{kurakin2016adversarial}, which we chose due to its simplicity, short runtime, and reliable results.
Then, to update $v$ (line 6), we use \emph{Projection} to project $v+v_{i}$ into an epsilon ball under the $L_{2}$ distance, thus ensuring that $\|v\|_2 \leq \epsilon$.
If the algorithm finds a TUAP where $TFR(f(X + v), y) \geq \delta$ (line 9), then it returns $v$. 
Otherwise the loop ends after $E$ iterations without finding a result, and the attacker should consider changing the constraints (i.e., $\delta$ or $\epsilon$).

\begin{algorithm}[t]
    \caption{Computation of targeted universal adversarial perturbation: for dataset $X$, target class $y$, alpha model $f$, maximal perturbation size $\epsilon$, minimal expected targeted fooling rate (TFR) $\delta$, and maximal number of iterations $E$.}
\begin{flushleft}
    \textbf{Input:} $X$, $y$, $f$, $\epsilon$, $\delta$, $E$.\\
    \textbf{Output:} a targeted universal adversarial perturbation $v$.
\end{flushleft}
\label{attackAlgo}
\begin{algorithmic}[1]
    \State initialize $v \gets 0$
    \For{$k=1$ up to $E$}
        \For{each batch $X_{B}$ in $X$}
            \If{$TFR(f(X_{B} + v), y) < \delta$}
                \State $v_{i} \gets \arg\min_{r}\|r\|_{2}\, \hspace{0.5cm} s.t.\ TFR(f(X_{B}+v+r), y) \geq \delta$
                 \State $v \gets Projection(v + v_{i}, \epsilon)$
            \EndIf
        \EndFor
        \If{$TFR(f(X + v), y) \geq \delta$}
            \State \Return $v$
        \EndIf
    \EndFor
\end{algorithmic}
\end{algorithm}

\section{Evaluation Setup}

\subsection{Dataset}
We use real intraday market data from the S\&P 500 index from Kaggle~\cite{snp500dataset}. 
For each stock, the dataset contains the open-high-low-close data at one-minute intervals between 11/9/2017-16/2/2018.
We define an input sample as a stream of 25 continuous one-minute records, where each record consists of the opening price, high price, low price, closing price, and the traded stock volume at the end of the minute.
The dataset is divided into a set for training the alpha models, a set for crafting TUAPs, and six test sets to evaluate the attack.
The data between 11/9/2017-1/1/2018 is used to train the alpha models.
To craft the TUAPs, 40 samples are uniformly sampled from each of the three trading days between 2/1/2018-4/1/2018 (a total of 120 samples).
The data between 5/1/2018-15/2/2018 (five trading days per week) is used to create six test sets: $T_1,...,T_6$.
For each week, we build a corresponding test set by uniformly sampling 100 samples that represent an increase of the stock price and an additional 100 that represent a decrease of the stock price.
For $1 \leq i \leq 6$, $T_i$ denotes the test set of the $i$'th week (i.e., we used the first week for $T_1$, etc.).
This sampling method ensures that our evaluation process is not influenced by imbalances in the data.

\subsection{Feature Extraction}
Before feeding the input into the models, we perform preprocessing on the raw data based on~\cite{arevalo2016high}.
However, while~\cite{arevalo2016high} only used the closing price of each minute, we aggregate five groups of five consecutive minutes, and for each group we extract the following features: the trend's indicator, standard deviation of the price, and average price.
The trend indicator is set to be the linear coefficient among the closing prices of five consecutive minutes. 
The features extracted from the sliding window of 25 minutes of the raw data are used to build one input sample for our alpha model.
Each sample is a vector of the following features: the last five pseudo-log returns, the last five standard deviations of price, the last five trend indicators, the last minute, and the last hour.
The pseudo-log return is calculated based on the average price of each five-minute group ${AVGp_{1},...,{AVGp_{5}}}$ and defined as $log(\frac{AVGp_{i}}{AVGp_{i-1}})$.
Thus, the preprocessing takes a sliding window of 25 minutes of the raw data and creates an input sample with 17 features.

\subsection{Models}
We use TensorFlow and Keras to implement three supervised alpha models which, for each processed sample, predict the stock's price movement at the end of the next five minutes. 
The models differ in terms of their architecture:
The DNN is a deep neural network with five hidden dense layers and a softmax layer; the CNN has a 1D convolution layer, two dense layers, and a softmax layer; and the RNN has two LSTM layers, a dense layer, and a softmax layer.
Since a model with high directional accuracy (DA) allows the user to develop a profitable investment strategy, the models were trained on the same data to maximize the DA. 
The models achieve 66.6\%-67.2\% and 65.6\%-68.3\% DA on the training and test sets respectively, promising results when compared to the results of other HFT alpha models~\cite{giacomel2015algorithmic,arevalo2016high,chenapplication}.
Since the models perform binary classification, a DA above 50\% can be used to build a profitable trading strategy.

\subsection{Evaluation}
For simplicity, we create TUAPs that force the three alpha models to predict that the stock price will increase.
For each model, we evaluate the TUAPs' performance using the six test sets ($T_1 - T_6$) with the following measurements: targeted fooling rate (TFR), untargeted fooling rate (UFR), and perturbation size. 
The TFR denotes the percentage of input samples that are classified as the target label and reflects the attacker's ability to control the AT system's decision regardless of the stock’s state. 
Although technical analysis is common in HFT, to the best of our knowledge, there are no known attacks against HFT AT systems that use technical indicators.
Therefore, to demonstrate the attack's added value, we compare the attack to random perturbations of the same size. 
However, measuring random perturbations with the TFR fails to reflect the unwanted prediction flips caused by random noise.
Thus, we also measure the UFR of the perturbations, which is defined as the percentage of input samples that were misclassified.
It is important to note that the goal of the TUAP is to cause all samples to be classified as the adversary's target, regardless of the true label.
Therefore, a successful attack is measured as the TFR, while the UFR measures the randomness effect of the perturbation on the prediction result.
The perturbation size is the average percentage of change in the stock’s closing price, which ensures that the perturbation remains undetected for each stock; 
hence a TUAP the size of a dollar is small for a stock that is $\$2000$ a share yet dramatic for a penny stock.
The relative size is also helpful for comparing the effects of different perturbations on different stocks.
Therefore, in this study, all of the perturbations are relative to the target stock price: $0.02\%$ of the price.
Our code is available at \url{https://github.com/nehemya/Algo-Trade-Adversarial-Examples}

\section{White-Box Attack}
In this experiment, we examine five stocks: Google (GOOG), Amazon (AMZN), BlackRock (BLK), IBM, and Apple (AAPL).
For each stock, we use Algorithm \ref{attackAlgo} to craft three TUAPs, one for every alpha model, and randomly sample three perturbations that are the same size as the TUAPs (i.e., $0.02\%$ of the stock's price).
Then, we evaluate the attack performance on six test sets $T_1 - T_6$.
Since the perturbations are trained on data from one point in time and evaluated on six unknown test sets from later periods of time, we expect that the performance of the TUAPs will gradually degrade as we move away in time from the time period of the training set; hence, the TUAP will achieve the highest TFR and UFR for $T_1$, when $T_6$ is the lowest.
Yet, the random perturbations are not expected to show any predictable behavior, and their effect on the classification result will probably not correlate to the time that has elapsed from the training set time. 

\begin{figure}[t]
    \centering
    \null\hfill
    \subfloat[DNN]{\label{fig:tfr-dnn}
        \includegraphics[width=0.32\textwidth]{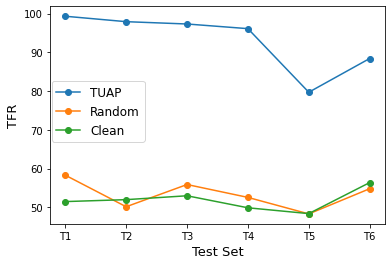}}\hfill
    \subfloat[CNN]{\label{fig:tfr-cnn}
        \includegraphics[width=0.32\textwidth]{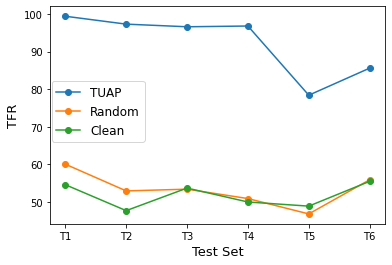}}\hfill
    \subfloat[RNN]{\label{fig:tfr-rnn}
        \includegraphics[width=0.32\textwidth]{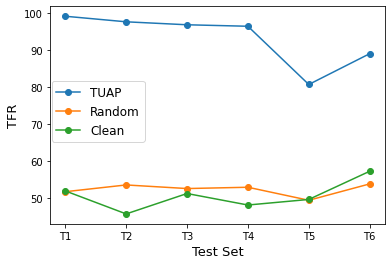}}\hfill\null
    \caption{  
        The mean TFR (percentage) of five stocks for the TUAP, random perturbation, and clean data.  
        The TFR is presented for each of the six test sets and the three models.
    }
    \label{fig:tfr}
\end{figure}

\begin{figure} [t]
    \centering
    \null\hfill
    \subfloat[DNN]{\label{fig:ufr-dnn}
        \includegraphics[width=0.32\linewidth]{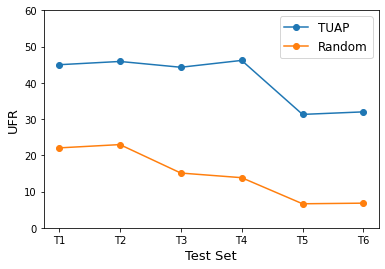}}\hfill
    \subfloat[CNN]{\label{fig:ufr-cnn}
        \includegraphics[width=0.32\linewidth]{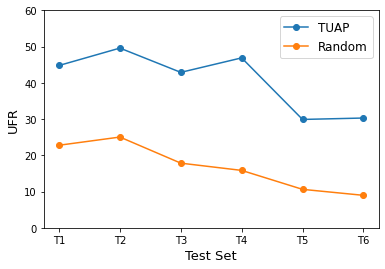}}\hfill
    \subfloat[RNN]{\label{fig:ufr-rnn}
        \includegraphics[width=0.32\linewidth]{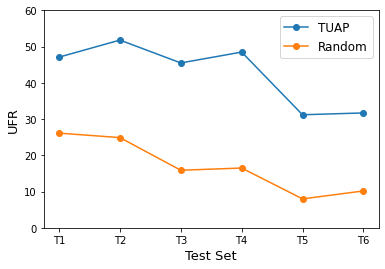}}\hfill\null
    
    \caption{
        The mean UFR (percentage) of five stocks for the TUAP and random perturbation.  
        The TFR is presented for each of the six test sets and the three models.
    }
    \label{fig:ufr}
\end{figure}

We examine the TFR for the TUAP and a random perturbation of the same size and compare them to the clean (original) results for each of the three alpha models.
These results for the random perturbation and clean data (see Fig.~\ref{fig:tfr}) indicate that the random noise does not have a major impact on any of the models.
On average, the random perturbations cause changes in the TFR that do not exceed 2\%, and thus the alpha models' prediction is not affected by them.
However, the TUAP creates dramatic changes in all of the alpha models' classification results, and the average TFR scores obtained by the TUAP are greater than $92\%$.
The results support the hypothesis that the attacker can use the TUAP to control the alpha model's prediction. 
To improve our understanding of the effect of the random perturbation, we also examine the UFR (see Fig.~\ref{fig:ufr}).
The results suggest that the TUAP causes a higher UFR than the random perturbation.
However, Fig.~\ref{fig:tfr} suggests that the classification flips caused by the TUAP are the result of input samples that have been pushed to be classified as the target label. 
Such effects are not demonstrated by the random perturbation.

As expected, while the effects of the random perturbations do not follow a certain pattern, the TUAP's TFR is highest for the earlier test sets (e.g., $T_1$) and gradually decreases for the later test sets.
However, an exception is found for $T_5$, which represents data from 2/2/2018-8/2/2018 and demonstrates poorer performance than any other test set, including the later test set $T_6$.
This exception may stem from several massive drops in the market that occurred during this time period~\cite{February2018Drop,stacey2018stock}. 
Although our datasets are balanced, each sample is based on 25 minutes of market data which reflect the drop in the market.
Thus, fooling an alpha model to predict an increase in a stock value in this week is a challenging task for a TUAP that was trained on normal market data.

Additionally, due to the heavy monitoring of the stock market, an attacker should consider the TUAP's size: a larger perturbation is more likely to fool the alpha model, but it also has a higher risk of being detected. 
Although there is a need to understand how the perturbation size affects the fooling rate, since we were able to achieve a high TFR by using TUAPs with a small size relative to the stock's price (i.e., $0.02\%$), we now examine the attack's effect on stocks in different price ranges, thus inspecting perturbations from similar absolute sizes (i.e., the number of dollars).
To do so, we define three price categories and randomly choose five stocks from each category: 
\begin{itemize}
    \item High ($\$900+$): GOOG, GOOGL, AMZN, ISRG, and PCLN.
    \item Medium ($\$400-\$650$): BLK, REGN, EQIX, AZO, and MTD.
    \item Low ($\$100-\$200$): ADP, UPS, IBM, AAPL, and AON.
\end{itemize}

As shown in Table \ref{tab:inter-cat}, the TUAPs fool the alpha models for stock data from all price categories examined, with a high TFR (greater than $89.55\%$). 
While the relative perturbation size is similar for all TUAPs, the effect on low-priced stocks is the greatest; the medium-priced stocks are less affected by the attack.
The results regarding the untargeted fooling rate support the findings from our previous experiment. 
We note that we examine the low-priced stocks in data from the S\&P 500 index, which contains only large companies.
Since the results indicate that targeting the low-priced stocks is the safer option for the attacker, we believe that future work should examine stocks with a price lower than 100\$.

\begin{table}[t]
\centering
    \setlength{\tabcolsep}{5pt}
    \begin{tabular}{l|cc|cccc}
        \toprule
        \thead[lb]{Stock Price \\ Category} &  \thead[b]{Average \\ Stock Price}   & \thead[b]{Average \\ TUAP Size} & \thead[b]{Clean \\ DA}  & \thead[b]{TFR \\ TUAP} & \thead[b]{UFR \\ TUAP} & \thead[b]{UFR \\ Random}  \\ 
        \midrule
        High     & \$1100     & \$0.266     & 66.65       & 91.35       & 41.41       & 15.64        \\ 
        Medium   & \$545      & \$0.152     & 68.01       & 89.55       & 39.58       & 15.96        \\ 
        Low      & \$141      & \$0.036     & 67.26       & 93.93       & 43.93       & 18.08        \\ 
        \bottomrule
    \end{tabular}
    \caption{
        Comparison of the TUAP and random perturbation effect on stocks at three price levels, including the average stock price and average absolute perturbation price.
        Each category is evaluated using the directional accuracy on clean data, the TFR and UFR on data with the TUAP, and the UFR on data with the random perturbation.
    }
\label{tab:inter-cat}
\end{table}

\section{Black-Box Attack}
Assuming the attacker has full knowledge of the target system is unrealistic in the real world where traders conceal this information.
Therefore, we evaluate the attack under black-box settings, where the attacker only has access to the model's input.
Since the attacker can manipulate the market stream, he/she can utilize the TUAP's transferability property to attack AT systems with an unknown alpha model architecture. 
Thus, the broker sends compromised market data targeting one AT system, which could affect other bots as well, and in turn, can potentially influence the entire market.
A transferable TUAP has a higher risk trading domain than other domains, because traders are protective of their AT systems' architecture, resulting in a lack of open-source AT implementations; this may encourage inexperienced traders to use similar alpha model architecture.

In this experiment, we craft a TUAP for each alpha model, which is then transferred to the other two models.
The results (see Table \ref{tab:transferability}) show that all of the TUAPs were highly transferable and sometimes achieved a TFR similar to that of the attacked model.
This likely stems from the models' preprocessing; although the alpha models have different learning model architectures, because they shared parts of the code, they had the same vulnerability. 
However, our results indicate that an attack from a malicious or compromised broker targeting a popular open-source implementation of an AT system could affect other bots using that software.
The presence of multiple compromised AT systems could cause a misrepresentation of the market's supply and demand, which might affect the behavior of AT systems that do not share that vulnerability.

\begin{table}[t]
\centering
    \null\hfill
    \subfloat[IBM]{
        \begin{tabularx}{0.47\textwidth} {
            c
            | >{\centering\arraybackslash}X
            | >{\centering\arraybackslash}X >{\centering\arraybackslash}X >{\centering\arraybackslash}X
            | >{\centering\arraybackslash}X}
            \toprule 
            \multicolumn{2}{c|}{Target} & DNN   & CNN   & RNN     & Size  \\ 
            \midrule
            \multirow{3}{*}{\rotatebox{90}{Source}}  & DNN   & -     & 95.04 & 95.00      & 0.027\\ 
                                                     & CNN   & 94.45 & -     & 94.75   & 0.028\\ 
                                                     & RNN   & 94.25 & 95.00    & -       & 0.026\\ 
            \bottomrule
        \end{tabularx}
    }\hfill
    \subfloat[AAPL]{
        \begin{tabularx}{0.47\textwidth} {
                c
                | >{\centering\arraybackslash}X
                | >{\centering\arraybackslash}X >{\centering\arraybackslash}X >{\centering\arraybackslash}X
                | >{\centering\arraybackslash}X}
            \toprule 
            \multicolumn{2}{c|}{Target} & DNN   & CNN   & RNN     & Size  \\ 
            \midrule
            \multirow{3}{*}{\rotatebox{90}{Source}}  & DNN  & -     & 93.7  & 93.66 & 0.024\\ 
                                                     & CNN  & 93.16 & -     & 93.41 & 0.022\\
                                                     & RNN  & 93.95 & 94.62 & -     & 0.024\\
            \bottomrule
        \end{tabularx}
    }\hfill\null
\caption{The transferability (TFR) of the TUAP between the three alpha models on (a) IBM and (b) Apple data.
The rows are the source (surrogate) models, the columns define the target (unknown) model, and the size denotes the relative TUAP size.
}
\label{tab:transferability}
\end{table}

\section{Mitigation}
While technology is advancing at a rapid pace, government regulation and enforcement systems are largely unaware of the risks that such advancement poses (e.g., the threat of adversarial examples).
Unlike known cyber attacks on the stock market which perform notable transactions (e.g., hack, pump and dump \cite{nakashima2007hack}), our attack performs small perturbations that can be concealed and considered a common error in the data.
The combination of an imperceptible attack and lack of knowledge about this threat allows attackers to exploit the market with minimal risk.
Therefore, we call on the finance community to raise awareness of the risks associated with the use of machine learning models in AT.
Since the traders using the AT systems will be the first entity to suffer from our attack, we examine several mitigation methods that can be used to protect AT bots.
In this section, we assume that the TUAP, its training set, and the percentage of perturbed samples in the test sets are known to the defender, and the attacker cannot tweak the attack to bypass the defense~\cite{carlini2017adversarial}.
Although it is an unrealistic assumption which gives a major edge to the defender, it allows us to discuss the challenges of protecting against adversarial perturbations in AT.

\begin{figure}[t]
    \centering
    \null\hfill
    \subfloat[kNN]{\label{fig:detector-knn}
        \includegraphics[width=0.4\linewidth]{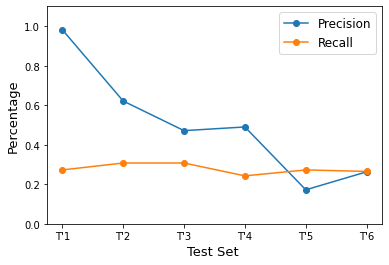}}\hfill
    \subfloat[ANN]{\label{fig:detector-ann}
        \includegraphics[width=0.4\linewidth]{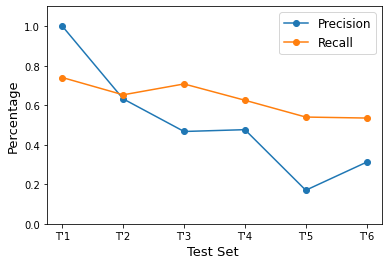}}
    \hfill\null
    \caption{
        The precision (blue) and recall (orange) of supervised detectors trained on the TUAP training set, with 10\% perturbed data: (a) kNN and (b) neural networks.
    }
    \label{fig:detector}
\end{figure}

A common mitigation approach involves the use of a detector to identify perturbed data and filter it out.
Due to the defender's unrealistic advantage, we can use the TUAP to build two simple classifiers, with k-nearest neighbors (kNN) and ANN architectures, to identify perturbed data.
We trained the detectors on the same training set used to craft the TUAP but added the perturbation to 10\% of the data.
Then, we created $T'_1,...,T'_6$ test sets, such that $T'_i$ is a combination of 10\% perturbed data (i.e., $T_i$) and 90\% benign data that was sampled from $i$th week.
Fig.~\ref{fig:detector} shows that while the kNN detector failed to detect the perturbed data, the ANN detector identified more samples but obtained a high false positive rate, which makes it unreliable.
A possible explanation for the results is that identifying adversarial perturbations requires the detector to model and predict normal behavior of the market, a challenging task that the entire AT system tries to perform.
Moreover, the performance of both detectors decreases as time passes; thus, the defender will have to retrain a new detector every few weeks.
This is unlikely, since in real life the defender would have to build the detector for unseen TUAPs without knowing their distribution in the market data. 
Additionally, a complex detector adds computational overhead to the AT system, which can make it irrelevant in HFT.
Therefore, some traders might prefer to risk being attacked in order to maintain an effective AT system.

\begin{figure}[t]
    \centering
    \null\hfill
    \subfloat[TUAP (TFR)]{\label{fig:adv_train-tfr}
        \includegraphics[width=0.4\linewidth]{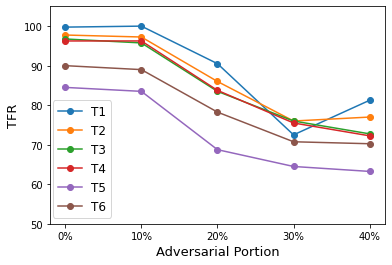}}
    \hfill
    \subfloat[Clean Data (DA)]{\label{fig:adv_train-da}
        \includegraphics[width=0.4\linewidth]{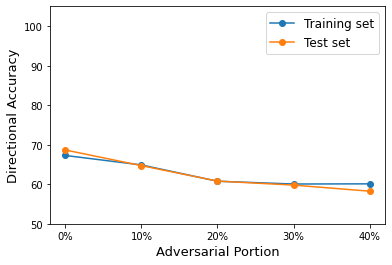}}
    \hfill\null
    \caption{
        The effects of adversarial retraining for different portions of adversarial examples in the retraining set on (a) the  perturbed evaluation sets (TFR), and (b) on the model's ability to predict the market for clean data (DA).
    }
    \label{fig:adv_train}
\end{figure}

An alternative approach is to train the model to correctly classifying both benign and perturbed data samples, which can be achieved by performing adversarial retraining~\cite{goodfellow2014explaining}: 
After training the alpha model, the defender creates perturbed data, which is labeled as the benign samples and then used to retrain the model.
By doing so, the defender avoids the computational overhead associated with an additional component to the AT system.
We used the attack's TUAP to perform adversarial retraining on the alpha model, and, as shown in Fig.~\ref{fig:adv_train}, the alpha model loses its ability to predict the market when the percentage of adversarial examples used for retraining increases.
When adversarial examples are 40\% of the retraining data, the TFR decreases from more than 90\% to around 70\%, and the directional accuracy drops from almost 70\% to a little less than 60\%, which indicates that the alpha model could make random predictions instead of learning from the training set.
Therefore, improving the model's robustness to adversarial examples is unsuitable for this domain.

Finally, we suggest a simple solution in which the trader collects the market data from several brokers, compares it, and filters out mismatches.
While the solution is trivial, it might be expensive for an individual trader who must pay multiple brokers for the market data.
Additionally, in HFT, the data might contain inaccuracies due to technical issues, or mismatches due to time differences between brokers operating around the world. 
As a result, traders may prefer to risk an attack over losing money or decreasing the AT system's performance.
However, regulatory authorities are not limited by the challenges discussed. 
They can perform offline analysis, compare the data from many brokers, and utilize complex detectors. 
Even if those methods are not used routinely, they can be applied when unexplained abnormal behavior is seen in the market.
Although securing the stock market may be challenging and no comprehensive solution currently exists, the war may already have begun , and efforts to defend against adversarial examples should start with increasing the finance community's awareness of the risks of using learning models in the stock market.

\section{Conclusions}
In this study, we suggested an attack that allows an attacker to control an AT system. 
By creating a TUAP and adding it to the market data stream in real time, an attacker can influence the alpha model's predictions.
Our results show that HFT systems are highly susceptible to adversarial inputs.
The use of a TUAP ensures that the attack can be performed in real life, where attackers cannot predict the market’s behavior. 
Adversarial perturbations are much stealthier than known manipulations against the stock market.
Since the size of all of the TUAPs created is 0.02\% of the stock price, the attack might not be flagged as such by monitoring systems and instead be viewed as a routine error in the data.
Our experiments also showed that TUAPs are transferable to different AT systems and that a perturbation targeting one alpha model can also fool another model.
Since the manipulated data in this study is sent by the broker, a TUAP added to the market data can change the behavior of multiple AT bots, and their actions may start a cascade effect, influencing other invulnerable systems and possibly the entire market. 
Given the lack of diversity in open-source bot implementations online, traders often use alpha models with similar architecture, a situation in which transferable perturbations can increase the potential damage, as shown in our study. 
Finally, we demonstrated potential mitigation methods against adversarial perturbations and discussed the concerning findings. 
Since many regulatory authorities and traders are unfamiliar with adversarial perturbation, we strongly suggest that the finance community examine this risk and take the steps required to protect financial markets from such threats.

%
%
%
\bibliographystyle{splncs04}
\bibliography{mybibliography}

\end{document}